\documentclass[rnote,oldversion]{aa}
\usepackage{graphicx}        

\begin{document}

   \title{ CADIS has seen the Virgo overdensity and parts of \\ the Monoceros
   and `Orphan' streams in retrospect  }

   \author{B. Fuchs\inst{1} \and S. Phleps\inst{2} 
   \and K. Meisenheimer\inst{3}}


   \institute{Astronomisches Rechen--Institut am Zentrum f\"ur Astronomie der 
              Universit\"at Heidelberg,
              M\"onchhofstrasse 12--14, 69120 Heidelberg, Germany
	      \and Max--Planck--Institut f\"ur extraterrestrische Physik,
               Giessenbachstrasse, 85748 Garching, Germany \and
	       Max--Planck--Institut f\"ur Astronomie, K\"onigstuhl 17, 69117
	       Heidelberg, Germany}

   \date{Received 2006; accepted}

   \abstract{
   We reanalyze deep star counts in five CADIS fields. The data are 
   presented as vertical density distributions of stars perpendicular to the
   Galactic plane. In three fields the profiles are consistent with each other,
   while in two fields significant overdensities of stars are found. The
   overdensity in one field can be associated with the Virgo overdensity which
   can be traced right into the disk of the Milky Way. Using this detection we
   estimate the mass of the Virgo overdensity and show that this is equivalent
   to the stellar content of a Local Group dwarf spheroidal galaxy. The 
   overdensity in the second field is more difficult to associate with a
   previously known overdensity. We suggest that it is related both to the 
   Monoceros stream and the recently discovered Orphan stream.
   
      }
		
   \mail{fuchs@ari.uni-heidelberg.de}
   
   \titlerunning{Reanalysis of CADIS data}
		
   \maketitle
%


In two previous papers (Phleps et al. 2000, 2005) we have studied the
distribution of stars in the Milky Way using deep star counts based on Calar 
Alto Deep Imaging Survey data. For this purpose we analyzed five fields which 
contained in total 1627 faint stars (15.5 $ < R < $ 23). We do not repeat here 
any details of our analysis, but give again in Table 1 the coordinates 
of the fields. At the time of writing the papers we were mainly interested in 
determining descriptive parameters of the density distributions
of the various components of the Milky Way 
such as the vertical scale heights of the thin and thick disks, respectively, 
their relative normalization and the parameters of the halo density law. We
realized, however, already then that, while the 1h, 16h, and 32h fields gave 
very consistent results, the 9h and to a lesser degree the 13h field lead 
to discrepant results in the sense that the vertical density distribution 
of the stars in these fields had a few kpc above the midplane a shallower slope 
than in the three other fields. With the advent of the Data Release 5 of the
Sloan Digital Sky Survey and its subsequent analyses by Juri\'c et al.~(2006) 
and Belokurov et al.~(2006a, b) overdensely populated parts of the density 
distribution of the stars in the Milky Way have been revealed in such rich 
detail that we can interpret now in retrospect also the results of the 13h
field in a consistent way, whereas the nature of the overdensity in 9h 
remains at present less clear. By pure chance two out of the five 
line--of--sights of the CADIS fields have crossed two separate of such 
overdense regions! Of course such few line--of--sights did not allow the 
identification of the excess densities as distinct isolated overdensely 
populated regions in the Milky Way. 

\begin{table}
\caption[]{Pointings of the CADIS fields}
         \label{poi}
\[
\begin{tabular}{rrrrr}
\hline
 \noalign{\smallskip}
 \hline
 \noalign{\smallskip}
  field & R.A. & Dec & l & b \\ 
       & \multicolumn{4}{c}{deg}\\
 \noalign{\smallskip}
 \hline
 \noalign{\smallskip}
 1\,h &	 27 &	  2 &   150 &	-59 \\
 9\,h & 138 &	 46 &	175 &	 45 \\
13\,h &	207 &	  6 &	335 &	 60 \\
16\,h & 246 &	 56 &	 85 &	 45 \\
23\,h & 349 &	 12 &	 90 &	-43 \\
 \noalign{\smallskip}
 \hline
 \end{tabular}
 \]
\end{table}

Such features in the density distribution of stars have attracted recently
great interest in the literature, because they almost certainly represent debris
of satellite galaxies which have fallen into the Milky Way and were then
disrupted. Particularly striking are the large filaments like the Sgr stream or 
the newly discovered `Orphan' stream (Ibata et al.~1997, Majewski et al.~2003, 
Belokurov et al.~2006a, b, Grillmair 2006) which are interpreted as tidal 
tails of dwarf galaxies presently in the process of being cannibalized. 
Obviously such accretion events played an important role in the formation 
history of the Milky Way.

In this {\em note} we present a reanalysis of the CADIS data, because in our 
view these data still contribute valuable information on the overdensities. 
In particular we can trace the overdensities right into the disk of the Milky
Way, which is not possible with the SDSS data. Moreover, we can provide 
estimates of the masses of the overdensities.

\begin{figure}
  \centering
  \includegraphics[width=8.7cm]{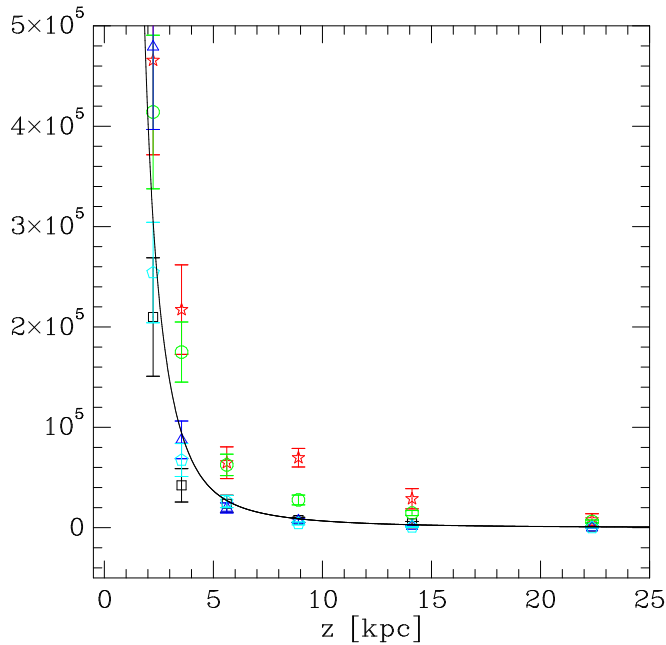}
  \caption[]{Number density distribution of stars [**/kpc$^3$]  
  perpendicular to the  Galactic plane derived from star counts in CADIS 
  fields. The five fields are coded as: 1h (squares), 9h (asterisks),
  13h (circles), 16h (triangles), 23h (pentagons). Only data above $z$ = 2 kpc 
  are shown (colour coded in the electronic version). The solid line is the 
  vertical density profile at the position of the Sun of a smooth Galaxy model 
  fitted to the data of the 1h, 16h and 23h fields.}
  \label{cadidat}
  \end{figure}

 The CADIS star counts were carried out along the lines--of--sight
whose coordinates are given in Table 1. In Fig.~1 we show the inferred number 
densities of stars not as function of heliocentric distance, but as function of
the distance from the midplane, i.e.~as vertical density profiles perpendicular 
to the Galactic plane. In this way the five fields are projected onto each 
other which allows a direct comparison of their density distributions. The 
distributions of the 1h and
23 h fields have been flipped up. As can be seen from Fig.~1 the distribution 
of stars is traced from the outer halo into the disk of the Milky Way. Data 
below $z$ = 2 kpc are not shown. The density distributions derived from the 
star counts in the 1h, 16h, and 23h fields, respectively, are within statistical
uncertainties consistent with each other. The inclinations of the
lines--of--sight of the CADIS fields relative to the vertical axis hardly affect
the shape of the density profiles at heights of more then 5 kpc above the
midplane. At lower heights part of the scatter among the data shown in Fig.~1
can be ascribed to the varying viewing directions of the CADIS fields. For the 
present purposes the 1h, 16h and 23h fields define a vertical reference profile.
Having realized this we have repeated the fit of the smooth Galaxy model of
Phleps et al.~(2005, Eqns.~3 and 5) using only these three fields. As in the
previous paper we adopt a vertical scale height of the thin disk of 283 pc. The
fit to the density distribution in the three reference fields leads to a 
slightly reduced vertical scale height of the thick disk of 900 pc, but to the 
same normalization of the local density of the thick disk at 4 percent of the 
total density at the midplane. In the case of a spherical halo model we find an 
index of the halo density law of $\alpha = 3.25\,\pm\,0.10$ and in the case of 
a halo flattened as $(c/a)=0.6$ an index of $\alpha= 2.69\,\pm\,0.09$. The 
density profile perpendicular to the Galactic plane ($b=90^\circ$) of the
smooth Galaxy model, which is shown as a solid line in Fig.~1, is in excellent 
agreement with the Galaxy model of Juri\'c et al.~(2006).

The 13h field shows a statistically significant excess density relative  
to the reference profile which can be traced from $z$ = 2 to 14 kpc with the
maximal deviation at about $z$ = 4 kpc. The heliocentric distances range 
from 2.3 to 16 kpc. The coordinates of this field point towards the fringe 
of the Virgo overdensity which is discussed in detail by Juri\'c et al.~(2006, 
cf.~their Fig.~24). Indeed, Fig.~1 can be directly compared with Fig.~22 
of Juri\'c et al.~(2006), where they delineate the overdensity by subtracting 
from the density distribution of stars observed in the meridional section of 
the Milky Way which contains the Virgo overdensity their smooth Galaxy model.
Above $z$ = 5 kpc, which corresponds to a galactocentric radius of $R$ = 5.5
kpc in the 13h field, the vertical profile of the overdensity found in the 
SDSS DR5 and the CADIS data, respectively, are absolutely consistent with each 
other. However, in the CADIS data it can be traced right down into the disk of 
the Milky Way, confirming the supposition of Juri\'c et al.~(2006) that this might
be the case. 

Moreover, both data sets have been analyzed as Hess diagrams. In 
their Fig.~24 Juri\'c et al.~(2006) show by subtracting the Hess diagram of a 
control field from the Hess diagram of the Virgo field that the excess 
population of stars in the Virgo field is primarily found in the blue branch of
the Hess diagram at $(g-r) \approx 0.4$, where halo stars are located. Precisely
the same is found in the Hess diagram of the 13h field shown in Fig.~2. There 
is a clear excess in the blue branch of the halo stars at $(b-r) \approx 0.1$ 
with respect to the smooth Galaxy model, which has been recalculated using the
parameters given above. We conclude from this discussion that the excess 
density in the 13h CADIS field can be identified as part of  the Virgo
overdensity.

\begin{figure}
  \centering
  \includegraphics[width=8.7cm]{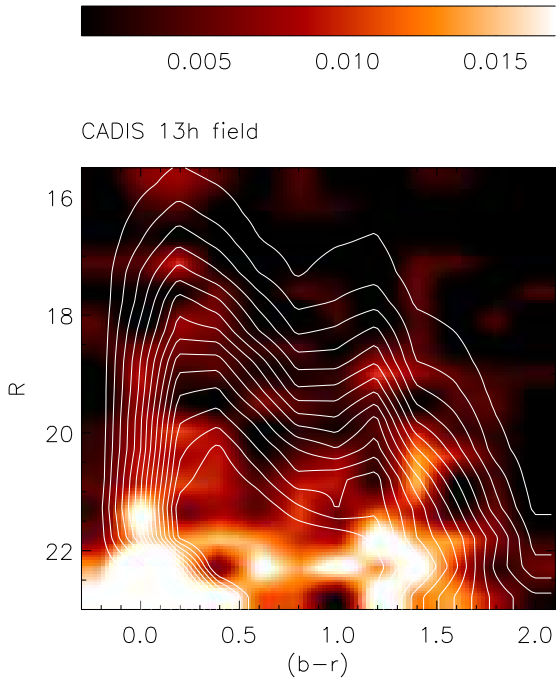}
  \caption[]{Hess diagram of the 13h field. The observed distribution of stars
  in the $R - (b-r)$ plane is colour coded according to the number density of
  stars per bin of size 0.1 $\times$ 0.1 in magnitude and colour, respectively. 
  The colour table is shown at the top. A smooth model of the Galaxy, which has
  been fitted to the reference fields, is shown as isodensity contours. The 
  equally spaced contour levels span the same range as the colour table.}
  \label{hess1}
  \end{figure}

Subtracting the reference profile from the observed density profile of the 13h 
field, both now reckoned along the line--of--sight, allows us to
determine the mass of the excess population of stars. We find that 276 out of
the 517 stars in the 13h field seem to belong to the density excess. Phleps et
al.~(2005) have identified in the CNS4 catalogue (Jahrei{\ss} \& Wielen 1997)
the analogues of the blue halo stars in the CADIS fields in the local volume
(Fuchs \& Jahrei{\ss} 1998).  They have shown that the extrapolation of the 
outer
halo density law towards the Galactic midplane agrees ideally with the number
density of stars with the same colours and absolute magnitudes in the local
sample. The average mass of these stars is 0.66 ${\mathcal{M}}_\odot$
and the average mass--to--light ratio is ${\mathcal{M}}/{\mathcal{L}_{\rm V}}
= 2.7 {\mathcal{M}}_\odot/{\mathcal{L}}_{{\rm V}\odot}$. 
Thus a number density of 10$^5$ stars per kpc$^3$ corresponds to a mass 
density of $6.6\,\cdot\,10^{-5}$ ${\mathcal{M}}_\odot\,pc^{-3}$. We estimate 
from the area preserving Lambert projection in Fig.~24 of Juri\'c et al.~(2006) 
that the Virgo overdensity subtends an area of 846 square degrees. One CADIS 
field has a size of 121 square arcminutes. If the mass of the excess population 
of stars in the 13h field is representative for the rest of the Virgo 
overdensity the latter contains a mass of $4.6\,\cdot\,10^6$ 
${\mathcal{M}}_\odot$. This is in our view a clear indication that the Virgo
overdensity is the relic of a shredded dwarf galaxy. Adopting the 
mass--to--light ratio of  ${2.7\,{\mathcal{M}}_\odot/
{\mathcal{L}}_{{\rm V}\odot}}$ found above the stars in the Virgo overdensity
would have a total luminosity of $1.7\,\cdot\,10^6$ 
${\mathcal{L}}_{{\rm V}\odot}$ or an absolute magnitude of M$_{\rm V}$ =
 -- 11 mag, which is quite typical for Local Group dwarf spheroidal galaxies. 
Also the inferred mass--to--light ratio is typical for the stellar populations 
of dwarf spheroidals (Mateo 1998).

\begin{figure}
  \centering
  \includegraphics[width=8.7cm]{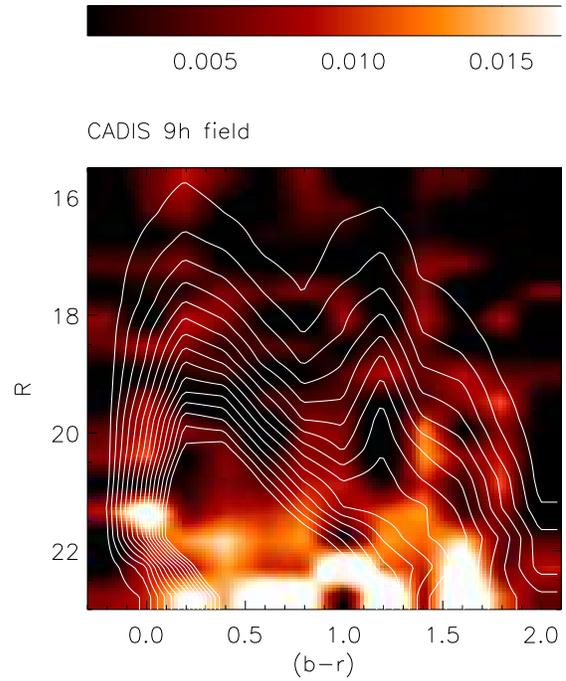}
  \caption[]{The same as Fig.~2 but for the 9h field.}
  \label{hess2}
  \end{figure}

The 9h field shows at distances between $z$ = 2 and 15 kpc above the midplane
an overdensity relative to the reference profile, which is even more 
pronounced than that in the 13h field. The maximal deviation is found around 
$z$ = 9 kpc. The corresponding heliocentric distances range from 3 to 21 kpc.
However, this overdensity is more difficult to associate confidently with one of
the overdensities seen in the SDSS data (Juri\'c et al.~2006, Belokurov et 
al.~2006a, b). We note that most of the density excess in the Hess diagram of
the 9h field shown in Fig.~3 is found in the colour range
$(b-r) \approx 0.5 - 1$ and falls right into the gap
between the two branches of the disk and halo stars, respectively. Judging from
the Hess diagrams presented by Juri\'c et al.~(2006) for the Virgo field and its
control field stars of such intermediate colour have been eliminated by the
colour cut $(g - r) < 0.4$ from the sample of Belokurov et al.~(2006a).  
In Fig.~1 of Belokurov (2006b) there is also the distribution of stars with
colours $0.4 < (g - r) \leq 0.6$ shown and we find the 9h field at the eastern 
fringe of the northern part of the Monoceros stream (Newberg et al.~2002) which
seems to be in this colour range much more extended than in the bluer 
$(g - r) < 0.4$ colours (Belokurov et al.~2006a). Similarly the Monoceros 
stream is seen in the data of Juri\'c et al.~(2006, their Fig.~9) mainly at 
slightly lower galactic latitudes than that of the 9h field. Its 
line--of--sight passes though through the 
inner fringe of the overdensity in anticenter direction in the panels of
Fig.~9 which show the density distribution of stars with colours $0.1 < 
(r - i) \leq 0.15$ at heights of 4 and 5 kpc above the midplane. Juri\'c et
al.~(2006) do study the distribution of redder stars, but cannot trace it beyond
a few kpc from the Sun.  Pe\~narrubia et al.~(2005) have modelled the Monoceros
stream by numerical simulations as a tidal stream. Their simulations  
show that part of the stream might be very well seen in the direction of the 9h 
field and at the heliocentric distances of the overdensity in the 9h field.
Thus this overdensity may be tentatively associated with the Monoceros 
stream. However, there is a further aspect of the interpretation of the
overdensity in the 9h field. The 9h field is positioned exactly on the 
galactocentric great circle on which the Orphan stream lies and which passes 
also through the high velocity cloud complex A (Belokurov et al.~2006a, b,
Wakker 2001). The position is roughly in the middle between the northern tip of
the Orphan stream and Complex A. Belokurov et al.~(2006b) have determined 
a heliocentric distance of the northern tip of the Orphan stream of 35$\pm$10
kpc, whereas the distance to Complex A is 10.1$\pm$0.9 kpc. This discrepancy 
can be resolved in a natural way if Complex A is on a different wrap around the
Galactic center as the Orphan stream. The density profile of the 9h field
gives the impression that there is an extra `hump' in the density excess of
stars relative to the reference fields which is centered around z
$\approx$ 9 kpc or a heliocentric distance of 13 kpc. It is tempting to
speculate that the `hump stars' are associated with the second wrap of
the Orphan stream. The width of the Orphan stream is estimated to be only a few
100 pc (Belokurov et al.~2006b). That would be consistent with a narrow feature
in the density profile of the overdensity. The fairly long elongated 
overdensity in the density profile of the 9h field must be then still ascribed
to the Monoceros stream.

In summary, we conclude that the Virgo overdensity has almost certainly been 
seen in the 13h CADIS field, although it was not recognized previously by us as
such. We have found in the 9h field another overdensity which is as significant
as the Virgo overdensity in the 13h field which we tentatively attribute to the
Monoceros and Orphan streams. Interestingly both features could be traced to 
distances less than 3 kpc from the Sun. If they are related to any of the star 
streams identified as fine structure in the phase space distribution function 
of stars in the solar neighbourhood (Helmi et al.~1999, Chiba \& Beers 2001, 
Navarro et al.~2004, Helmi et al.~2006, Arifyanto \& Fuchs 2006) is at present 
unclear.

\acknowledgements{We thank Vasily Belokurov, Wyn Evans and 
Hans--Walter Rix for very helpful discussions.}

{}

\end{document}